\documentclass[12pt,notitlepage]{article}
\usepackage{amssymb}

\usepackage{graphicx}
\usepackage{amsmath}

\input{tcilatex}

\begin{document}

\begin{center}
{\LARGE VACUUM INSTABILITY}

Paul S. Wesson
\end{center}

Department of Physics, University of Waterloo, Waterloo, Ontario \ N2L
3G1, Canada

\underline{PACS}: 0420, 0450, 1190, 9530

arXiv: gr-qc/0407038 v.1 (9 Jul 2004), v.2 (4 Feb 2005)


\underline{Abstract}: Following fresh attempts to resolve the problem of the
energy density of the vacuum, we reconsider the case where the cosmological
constant is derived from a higher-dimensional version of general relativity,
and interpret the gauge-dependence of $\Lambda $ as a dynamical effect. \
This leads to a relation between the change in $\Lambda $ \ and the line
element (action) which is independent of gauge choices and fundamental
constants: $d\Lambda ds^{2}=-6$. \ This implies that the (classical) vacuum
is unstable, with implications for particle production.

Correspondence: mail to P.S. Wesson, fax to (519)746-8115.\newpage

\section{\protect\underline{Introduction}}

The cosmological-constant problem has a long history, and while there are
many possible resolutions, none has gained widespread acceptance. \ In
classical general relativity, the energy density and pressure of the vacuum
obey $\rho c^{2}=-p=\Lambda c^{4} / 8\pi G$, where $c$ is the speed of light
and $G$ is the gravitational constant. \ The astrophysically-determined
value of $\Lambda $, for the present epoch at least, is small. \ But in
quantum theory, the vacuum (or zero-point) energies associated with particle
interactions lead to a value of $\Lambda $ which is big. \ The discrepancy
may be large as 10$^{120}$. \ Padmanabhan has recently reviewed this
problem, and outlined a resolution wherein the classical value of $\Lambda $
is essentially the statistical one ``left over'' from numerous stronger
interactions described by quantum field theory [1, 2]. \ Mashhoon and Wesson
have recently reconsidered the case where $\Lambda $ in four-dimensional
general relativity is derived from a five-dimensional formalism, such as
membrane or induced-matter theory, and found that the classical value can
depend on the size of the extra coordinate [3, 4]. \ In what follows, we
will present a short analysis that has something in common with both of the
aforementioned approaches, and calculate the change in the 4D cosmological
"constant" due to a change in the size of the fifth coordinate. \ 
We will work with a specific
gauge in order to get an answer, but the latter will turn out to be
independent of choices of gauge and fundamental constants. \ The equation
concerned is $d\Lambda ds^{2}=-6$. \ This relates the change in the
cosmological \ constant (or energy density of the vacuum) to the change in
the interval (or action for a particle of unit mass). \ It implies that the
classical vacuum is unstable. \ This invites application to a Dirac-like
model, where fluctuations in a vacuum field are balanced by the production of
massive particles.

The present account is brief and exploratory. \ But we believe that this
approach is worth pursuing, since while it is somewhat phenomenological it
has numerous applications, particularly to cosmology.

\section{\protect\underline{Relations for Vacuum Instability}}

In this section, we make use of technical results derived from 5D field
theory. \ This in general describes the classical fields associated with the
spin-2 graviton (Einstein gravity), the spin-1 photon (Maxwell
electromagnetism) and a spin-0 scalaron (Higgs-type mass field). \ It is the
basic extension of 4D general relativity, and is commonly regarded as the
low-energy limit of 10, 11 and 26 D (etc.) theories which may lead to a
grand unification of all of the known interactions [5]. \ There are
currently two versions of 5D field theory in vogue, namely membrane theory
[6] and induced-matter theory [7]. \ Both make essential use of a
non-compact extra dimension (which we label $x^{4}=l$, where the spacetime
coordinates are $x^{\alpha }$ with $\alpha =0$, 123; we temporarily absorb $c$ and $G$
via a choice of units which renders them unity). \ Physically, membrane
theory allows gravity to propagate freely (into the ``bulk''), whereas other
interactions are confined to a singular hypersurface (the ``brane''), thus
giving insight into the hierarchy problem and the masses of particles. \
Alternatively, induced-matter (or space-time-matter) theory places no
restrictions on the dynamics other than those which follow from solving the
geodesic equation, using Campbell's theorem to go from 5D to 4D and giving
an account of matter in terms of pure geometry. \ Mathematically, the two
theories are equivalent: (a) The field equations contain the same
information (the non-linear terms associated with the brane are contained as vector
components of the complete energy-momentum tensor). (b) Both theories involve an extra
force associated with the extra dimension (the discontinuities across the
brane balance, and reproduce the acceleration derived from the 5D geodesic).
\ (c) In either approach a massless particle in 5D can be viewed as a
massive particle in 4D (the photon is unique, being confined to the
hypersurface $l=0$). \ For technical details on these three points, the
reader is referred to [8], [9] and [10] respectively. \ We will use some of
the relevant technical results below, but our starting point will be the
recent analysis of the gauge-dependence of the cosmological constant $%
\Lambda $ referred to above [3]. \ This employs the ``canonical'' gauge of
induced-matter theory, which via a quadratic in $l$ defines a coordinate
system analogous to the synchronous one of standard cosmology, and leads to
a ready comparison with the usual action and masses of particles. \
Alternatively, there is the ``warp'' metric of membrane theory, which via an
exponential in $l$ defines a coordinate system similar to that used in
deSitter cosmology, and weakens gravity away from the brane and leads to an
explanation of why particles have masses less than the Plank value. \ These
two gauges are both valid, but using the former we will obtain a result
which is independent of either.

Consider then a 5D line element which contains the 4D one and depends on a
constant length $\left( L\right) $:%
\begin{eqnarray}
dS^{2} &=&\frac{l^{2}}{L^{2}}ds^{2}-dl^{2} \\
ds^{2} &\equiv &g_{\alpha \beta }\left( x^{\gamma },l\right) dx^{\alpha
}dx^{\beta }\;\;\;\;\;.
\end{eqnarray}%
Here the metric tensor can in principal depend on $x^{4}=l$, in which case
(1) is still general, because it uses the 5 available degrees of coordinate
freedom to remove the electromagnetic potentials $\left( g_{04}=0\right) $
and flatten the scalar potential $\left( g_{44}=-1\right) $, but imposes no
further constraints. \ For (1), numerous solutions are known of the
apparently-empty field equations, which in terms of the 5D Ricci tensor are%
\begin{equation}
R_{AB}=0\;\;\;\;\;\;\;\left( A,B=0,123,4\right) \;\;\;\;\;.
\end{equation}%
Using Campbell's theorem [11], it can be shown that these equations always
contain the equations of 4D general relativity, which in terms of the
Einstein tensor and an effective energy-momentum tensor are%
\begin{equation}
G_{\alpha \beta }=8\pi T_{\alpha \beta }\left( \alpha ,\beta =0,123\right)
\;\;\;\;\;.
\end{equation}%
In these, the energy density of the vacuum is nowadays frequently taken to
be implicit in $T_{\alpha \beta }$. \ But if it is taken to be explicit and
measured by $\Lambda $, then in the absence of ordinary matter the field
equations in terms of the 4D Ricci tensor are just%
\begin{equation}
R_{\alpha \beta }=\Lambda g_{\alpha \beta }\left( \alpha ,\beta
=0,123\right) \;\;\;\;\;.
\end{equation}%
These equations for metric (1) identify the length scale in the latter via 
\begin{equation}
\Lambda =3/ L^{2}\;\;\;\;\;.
\end{equation}%
This and the preceding results are by now well known [7]. \ However, it was
shown recently [3] that the gauge transformation $l\rightarrow \left(
l-l_{0}\right) $ in metrics of type (1) leads to a change in $\Lambda $ of
(6) to 
\begin{equation}
\Lambda =\frac{3}{L^{2}}\left( \frac{l}{l-l_{0}}\right) ^{2}\;\;\;\;\;.
\end{equation}%
This result is mathematically simple but physically profound. \ It indicates
that the 4D cosmological ``constant'', determined by (5), can diverge as one
approaches a 5D state $\left( l\rightarrow l_{0}\right) $ determined by (7).
\ The latter equation was arrived at by tedious algebra, and holds when the
4D part of the metric (1) has the conformally-flat or deSitter form $%
g_{\alpha \beta }\left( x^{\gamma },l\right) =f\left( x^{\gamma },l\right)
\eta _{\alpha \beta }$ [3]. \ This is a special case of the general
situation, that in non-compact 5D field theory the form of 4D quantities can
change under coordinate transformations that depend on $x^{4}=l$. \ An
alternative and instructive way to appreciate this kind of behaviour is as
follows:

A corollary of Campbell's theorem is that any solution of the source-free 5D
field equations $R_{\alpha \beta }=0$ with metric (1) can be written as a
solution of the empty 4D field equations $R_{\alpha \beta }=\Lambda $$%
g_{\alpha \beta }$ with $\Lambda =3/ L^{2}.$ \ Therefore, since the
equations are covariant, the same must hold for any gauge transformation
which leaves the form of the metric intact. \ For (1) with $l\rightarrow
\left( l-l_{0}\right) $, the fifth part of (1) is unchanged, while the
prefactor on the 4D part changes from $l^{2}/ L^{2}$ to $\left(
l-l_{0}\right) ^{2}/ L^{2}=\left( l^{2}/ L^{2}\right) \left[
\left( l-l_{0}\right) / l\right] ^{2}$. \ Let us write $\overline{g}%
_{\alpha \beta }=\left[ \left( l-l_{0}\right) / l\right] ^{2}g_{\alpha
\beta }$. \ Then the field equations hold with $\overline{R}_{\alpha \beta }=%
\overline{R}_{\alpha \beta }\left( \overline{g}_{\alpha \beta }\right) $ and 
$\overline{\Lambda }=\left( 3/ L^{2}\right) l^{2}/ \left(
l-l_{0}\right) ^{2}$. \ This is the same as (7). \ Put another way: a
translation along the fifth dimension necessarily changes the
four-dimensional cosmological ``constant''.

We now proceed to analyse the instability inherent in (7) by adding a series
of mathematical and physical conditions to the problem.

Firstly, let us take derivatives of (7), to obtain $d\Lambda = -(6/L^{2})(l-l_{0})^{-3}ll_{0}dl$. \
We are mainly interested in the region near $l=l_{0}$, where the energy density 
$\Lambda = \Lambda(l)$ is changing rapidly but smoothly (the change in the opposite
regime leads to the same relation, but less by a factor of 2).  \ Putting $dl=l-l_{0}$ for 
the change in the extra coordinate, we obtain
\begin{equation}
d\Lambda dl^{2}=-6l^{2}/L^{2}\;\;\;\;\;.
\end{equation}%
This is an alternative form of the instability inherent in (7) near to its divergence. 

Secondly, let us assume that the instability has a dynamical origin, and that the $l$-path involved
is part of a null 5D geodesic as in other work [10].  \ Then by (1) with $dS^{2}=0$, we have 
$l=l_{0}e^{\pm s/ L}$. \ We take the upper sign as elsewhere [3], which means that the path
drifts slowly away from the $l=l_{0}$ hypersurface. \ [The constant $L$ is large because
the current astrophysical value of $\Lambda$ is small, the two being inversely related by
(6). \ It should be noted that if we reverse the sign of the last term in (1), $\Lambda$ changes
sign and the path oscillates around $l_{0}$.] \  The noted path implies $dl/ l=ds/ L$, 
which in (8) yields
\begin{equation}
d\Lambda ds^{2}=-6\;\;\;\;\;.
\end{equation}%
This is remarkable, in that it contains no reference to $x^{4}=l$ and is
homogeneous in its physical dimensions (units), with no reference to
fundamental constants. \ That is, it is gauge and scale invariant. \ [An alternative
derivation of (9) may be made by using the expression for $\Lambda = \Lambda(s)$
found in ref. 3, equation (24) and noting that $s$ is measured from where $\Lambda$
diverges at the big bang.] \ Again (9) confirms the instability, since 
$d\Lambda \rightarrow \infty $ \ for $ds\rightarrow 0$. \ This behaviour 
can be put into better physical perspective by
recalling that the action for a particle of rest mass $m$ in 4D dynamics is
usually defined as $I=\int mds$. \ So (9) can be interpreted as a change in
the energy density of the vacuum for a particle of unit mass which changes
its action.

Thirdly, let us assume that the action is quantized. \ In most
higher-dimensional theories, the rest mass of a particle can change as it
pursues its 4D path, so $m=m\left( s\right) $ [9]. \ But irrespective of
this, we have with units restored that $dI=mcds=h$ where $h$ is Planck's
constant. \ Then (9) gives 
\begin{equation}
d\Lambda =-6\left( \frac{mc}{h}\right) ^{2}\;\;\;\;\;.
\end{equation}%
This says that a change in the energy density of the vacuum is related to
the square of the mass of a particle. \ The implication is clearly that the
vacuum (with an energy density proportional to $\Lambda $) gives up energy
which corresponds to a particle (with rest mass $m$). \ The precise fashion
in which this occurs cannot be investigated using the phenomenological
relation (10). \ However, the situation is similar to the old Dirac theory,
in which a positron is regarded as a hole created in an underlying sea of
energy. \ Another way of interpreting (10) involves geometry. \ Globally,
the vacuum is a sea of energy which curves spacetime,\ the
gravitationally-defined ``radius of curvature'' being related to $L=\sqrt{%
3/\Lambda}$ (see above). \ Locally, a perturbation in the vacuum
corresponds to a change in the curvature; and (10) in this picture says that
the change is related to the Compton wavelength $\left( h/ mc\right) $
of the particle. \ It is interesting to note that relations like (10) have
appeared previously in the literature [12]. \ Their rationale is to give a
semi-classical account of the origin of mass, a problem whose analog in
quantum theory involves the Higgs mechanism. \ At present, we are not sure
how to incorporate the symmetries manifested by particles into
higher-dimensional field theory. \ But (10) is a simple relation which is
compatible with other more detailed approaches.

\section{\protect\underline{Conclusion}}

When the cosmological ``constant'' $\Lambda $ \ as measured in a 4D
spacetime with proper time (action) $s$ is derived from a higher-dimensional
model, dynamical changes in these parameters are related by (9): $d\Lambda
ds^{2}=-6$. \ This is free of fundamental constants and other parameters
involving the choice of higher-dimensional gauge. \ It indicates that the
(classical) vacuum is unstable to spacetime changes. \ If the latter are
quantized, we obtain (10): this connects a change in the energy density of
the vacuum (as measured by $\Lambda $) to particle mass, in a way
reminiscent of the Dirac theory of the positron. \ Relations (9) and (10)
are phenomenological, insofar as they are derived in the context of 5D
(membrane and induced-matter) theory, without high-energy corrections from
other dimensions or input from models of particle interactions. \ In this
sense, they are like the relations of classical thermodynamics, which
however provide reasonable approximations without knowledge of the
underlying atomic physics. \ Both relations are compatible with recent
research on the cosmological-``constant'' problem and the nature of mass
[1-4, 7-12]. \ We are of the opinion that they provide a basis for more
detailed work, notably in the areas of particle production and
cosmology.\newpage 

\underline{{\LARGE Acknowledgements}}

This work was based on previous collaborations with B. Mashhoon and other people. \ It was
supported by N.S.E.R.C. and other agencies.

\underline{{\LARGE References}}

\begin{enumerate}
\item T. Padmanabhan, Phys. Rep. \underline{380}, 235 (2003).

\item T. Padmanabhan, Class. Quant. Grav. \underline{19}, 5387 (2002).

\item B. Mashhoon, P.S. Wesson, Class. Quant. Grav., \underline{21}, 3611 (2004).

\item B. Mashhoon, H. Liu, P.S. Wesson., Phys. Lett B\underline{331}, 305
(1994).

\item P. West, Introduction to Supersymmetry and Supergravity (World
Scientific, Singapore, 1986). \ M.B. Green, J.H. Schwarz, E. Witten,
Superstring Theory (Cambridge Un. Press, Cambridge, 1987).

\item L. Randall, R. Sundrum, Mod. Phys. Lett. A\underline{13}, 2807 (1998).
\ N. Arkani-Hamed, S. Dimopoulos, G.R.\ Dvali, Phys. Lett. B\underline{429},
263 (1998). \ R. Maartens, gr-qc/0312059 (2003).

\item P.S. Wesson, Space-Time-Matter (World Scientific, Singapore, 1999).

\item J. Ponce de Leon, Mod. Phys. Lett. A\underline{16}, 2291 (2001). \ J.
Ponce de Leon, Int. J. Mod. Phys. D\underline{11}, 1355 (2002).

\item B. Mashhoon, P.S. Wesson, H. Liu, Gen. Rel. Grav. \underline{30}, 555
(1998). \ P.S. Wesson, B. Mashhoon, H. Liu, W.N. Sajko, Phys. Lett. B%
\underline{456}, 34 (1999). \ D. Youm, Phys. Rev. D\underline{62}, 084002
(2000). \ J. Ponce de Leon, Grav. Cos. \underline{8}, 272 (2002). \ J. Ponce
de Leon, Int. J. Mod. Phys. D\underline{12}, 757 (2003). \ J. Ponce de Leon,
J. Gen. Rel. Grav. \underline{36}, 1335 (2004).

\item S.S. Seahra, P.S. Wesson, Gen. Rel. Grav. \underline{33}, 1731 (2001).
\ D. Youm, Mod. Phys. Lett. A\underline{16}, 2371 (2001).

\item J. Campbell, A Course on Differential Geometry (Clarendon Press,
Oxford, 1926). \ S.S. Seahra, P.S. Wesson, Class. Quant. Grav. \underline{20}%
, 1321 (2003).

\item E.A. Matute, Class. Quant. Grav. \underline{14}, 2771 (1997). H. Liu,
P.S. Wesson, Int. J. Mod. Phys. D\underline{7}, 737 (1998). F. Mansouri,
hep-th/0203150 (2002). P.S. Wesson, Class. Quant. Grav. \underline{19}, 2825
(2002).
\end{enumerate}

\end{document}